\documentclass[twocolumn,iop,floatfix,revtex4]{openjournal}
\usepackage[utf8]{inputenc}

\usepackage[british]{babel}

\usepackage{amsmath}
\usepackage{amssymb}
\usepackage{graphicx}
\usepackage{bm}
\usepackage{natbib}
\usepackage[colorlinks,allcolors=blue]{hyperref}
\usepackage{float}

\begin{document}
\journalinfo{The Open Journal of Astrophysics}
\submitted{Version January 16, 2024}

\shorttitle{Minor Merger Dual AGN}
\shortauthors{Mićić et al.}

\title{SDSS J125417.98+274004.6: An X-ray Detected Minor Merger Dual AGN}

\author{Marko Mićić $^{1\star}$}
\author{Brenna N. Wells $^{1}$}
\author{Olivia J. Holmes $^{1}$}
\author{Jimmy A. Irwin $^{1}$}
\affiliation{$^1$ Department of Physics and Astronomy, The University of Alabama, Tuscaloosa, AL 35401, USA}
\thanks{$^\star$ Corresponding author: \nolinkurl{mmicic@crimson.ua.edu}}

\begin{abstract}
In this paper, we present the discovery of a dual AGN in a $\sim$11:1 minor merger between the galaxy SDSS J125417.98+274004.6 and its unnamed dwarf satellite. We calculated stellar masses of the primary and secondary galaxy to be 2.5$\times$10$^{10}$M$_{\odot}$ and 2.2$\times$10$^{9}$M$_{\odot}$, respectively. We used archival Chandra X-ray observations to assess the presence of AGN. We found that both AGN have comparable luminosities of $\sim$2$\times$10$^{42}$ erg s$^{-1}$, with the secondary AGN being more likely to be the dominant one. The galaxies are in the early stages of the merger and are connected by a tidal bridge. Previous works suggest that the secondary AGN should experience a brief but intensive period of Eddington-limit approaching accretion during the early stages of the merger. During the merger, the secondary black hole can increase its mass by a factor of ten. SDSS J125417.98+274004.6 is the first known dual AGN in an early-stage minor merger with a comparably or more luminous secondary AGN. As such, it will be of great value for future studies of merger-triggered accretion and black hole growth mechanisms.
\end{abstract}


\maketitle

\section{Introduction}
\label{sec1}
Galaxy mergers are major drivers of galaxy evolution. They lead to changes in morphology (e.g.,  \cite{1972ApJ...178..623T,2008ApJ...672..177L}), color and chemical composition (e.g., \cite{2008ApJ...681..232L,10.1093/mnras/stad1503}), can affect star formation (e.g., \cite{Ellison_2008,10.1111/j.1365-2966.2012.21749.x,10.1093/mnrasl/slt136}), increase the stellar mass of the progenitor galaxies, and trigger the supermassive black hole accretion (e.g., \cite{refId0gao}). Since almost every galaxy is expected to harbor a supermassive black hole in its nucleus, galaxy mergers are a logical precursor to dual supermassive black hole systems. If during the merger both black holes turn active, a dual AGN will be formed. Dual AGN are of great interest for various astrophysical topics because they represent unique laboratories for studying the physics at the intersection of galaxy interactions, star formation, AGN activity, and gravitational waves.\newline \indent The connection between the occurrence of AGN and dual AGN and major mergers, involving two large galaxies with comparable masses, has been studied extensively. It is well established from the observational point of view that major mergers trigger the most luminous AGN and dual AGN \citep{2012ApJ...758L..39T,Comerford_2015,Donley_2018}. \cite{Koss_2012} also found that the dual AGN are more likely to occur in major mergers and that the X-ray luminosity of both AGN tends to increase with the decrease of galaxy separation, solidifying the merger-related accretion hypothesis. This is in accordance with the general picture in which major mergers cause disc instabilities, providing the fuel for the central black hole accretion \citep{2009Natur.457..451D}. Recent advancements utilizing the GAIA data provide an opportunity to detect previously inaccessible high-redshift sub-arcsecond separated dual AGN \citep{Mannucci2022}. Most of these dual AGN are powered by black holes with similar masses, implying they are relics of galaxy major mergers, offering new avenues to study the physics of in-spiraling pairs of supermassive black holes. On the other hand, dual AGN in mergers involving one or two dwarf galaxies are usually overlooked and understudied. \newline \indent Since every large galaxy is associated with multiple dwarf satellites, minor mergers, involving galaxies with disproportional masses (mass ratio at least 3:1), are much more common than major mergers. However, dual AGN in minor mergers are exceptionally rare, with only a handful of known objects \citep{Koss_2012,Comerford_2015,Secrest_2017,Liu_2018}. Various hydrodynamical simulations of dual AGN in minor mergers have been carried out \citep{Callegari_2011, VanWassenhove_2012,cap10.1093/mnras/stu2500}. They all found that during minor mergers, the primary AGN is mostly unaffected, but the secondary galaxy always responds strongly to the interaction. The secondary AGN grows significantly faster than the primary AGN, and the growth happens in two main phases. During the early stages of the merger, the secondary AGN is expected to accrete at high Eddington rates. However, ram pressure striping eventually removes all the gas from the secondary galaxy and the accretion ceases. The secondary AGN will accrete again once it enters the disk of the primary galaxy and starts consuming the gas reservoir. During this process, the secondary black hole can increase its mass by a factor of ten, but most of the time it is unable to accrete efficiently due to tidal forces. Furthermore, the activity of the primary and secondary black holes for most of the merger is not simultaneous, explaining the observational scarcity of minor merger dual AGN. As a result, the accretion physics and triggering mechanisms of minor merger dual AGN are not fully understood. \newline \indent Minor mergers in which the secondary AGN has comparable or higher luminosity than the primary AGN are rare. Was49 is an example of a late-stage minor merger, with a mass ratio between 1:7 and 1:15, where the dominant AGN resides in the smaller galaxy, the only known object of this nature \citep{Secrest_2017}, and as such represents a unique system for studies of interaction-triggered accretion onto secondary black holes. Some other notable works important for understanding the activity of AGN in minor mergers include discoveries of single AGN residing in dwarf galaxies undergoing mergers with their more massive companion. One such example is NGC 3341 \citep{b10.1093/mnras/stt1459}, a triple merger between a large disk galaxy, and two of its dwarf satellites, where a single AGN resides in one of the dwarfs. Some other works discovered intermediate-mass black hole candidates undergoing tidal disruption events, residing in dwarf galaxies that are being actively cannibalized by their larger neighbors (e.g., \cite{Lin2017}).
\newline \indent 
In this paper, we report the discovery of a new X-ray-detected dual AGN residing in an apparent early-stage minor merger of the galaxy SDSS J125417.98+274004.6 (SDSS J12, hereafter) with its smaller, unnamed companion (SDSS J12b, hereafter). The secondary AGN has a comparable X-ray luminosity to the primary AGN, making this the second such discovery, and the first in an early stage of the merger. \newline \indent
In Section 2 we describe the methodology and data reduction; in Section 3 we present our results; in Section 4 we discuss the implications of our findings. Throughout the paper, we assume the $\Lambda$CDM cosmology with parameters H$_0$=69.6 km s$^{-1}$ Mpc$^{-1}$, $\Omega_{\Lambda}$=0.71, $\Omega_{m}$=0.29. All errors are reported at the 1$\sigma$ level unless specified otherwise.
\begin{table}
	\centering
	\caption{Summary of the Chandra ACIS-I observations used in this paper.}
	\label{tab1:example_table}
	\begin{tabular}{lccc} 
		\hline
		Obs. ID & t$_{\textup{exp}}$ & Date & Off-axis angle\\
            & (ksec)& & (arcmin)\\
		\hline
		15023 & 43.5 & 2014-03-15 & 2.5\\
		15024 & 19.8 & 2014-03-16 & 2.6\\
		16599 & 28.7 & 2014-03-13 & 2.6\\
            16600  &96.8&2014-03-11&2.6\\      
		\hline
        \end{tabular}
\end{table}
\section{Methodology, Data, and Analysis}
\subsection{Target selection}
This discovery is part of a larger effort aimed at detecting dual AGN in previously unknown galaxy mergers. We first surveyed deep (t$_{exp}>$100 ksec) Chandra archival observations, searching for pairs of X-ray sources, separated by no more than 5$\arcsec$. We constructed a sample of 452 potential dual X-ray sources. The majority of potential pairs from this sample reside in galaxies with no available spectroscopic redshift or poorly constrained photometric redshift. This limitation makes calculating the physical distance between the two galaxies and assessing the stage of interaction impossible. We overcame this problem by using archival optical observations, mostly from the CFHT and DESI Legacy Survey data. Interacting galaxies often develop a variety of tidal features, such as tidal bridges, tails, and shells, and we resorted to detecting the existence of tidal debris around the galaxies hosting the potential X-ray sources as an indicator of an ongoing merger. A similar approach has been used in literature (e.g., \cite{2020AJ....159..103K}). We found that the overwhelming majority of our potential dual X-ray sources reside in galaxies that are chance-aligned background/foreground objects and not bona fide pairs, or in galaxies in which tidal debris is too faint to be detected. However, for a small subset of three potential pairs, the host galaxies exhibit clear tidal features and provide conclusive evidence for the galaxy association. Two of them are intriguing candidates for faint dwarf-dwarf mergers \citep{Mićić_2023}, while the third one, the subject of this paper, is an apparent minor merger dual AGN.
\subsection{X-ray and optical data} 
\begin{figure}[]
	\includegraphics[width=\columnwidth,height=7cm]{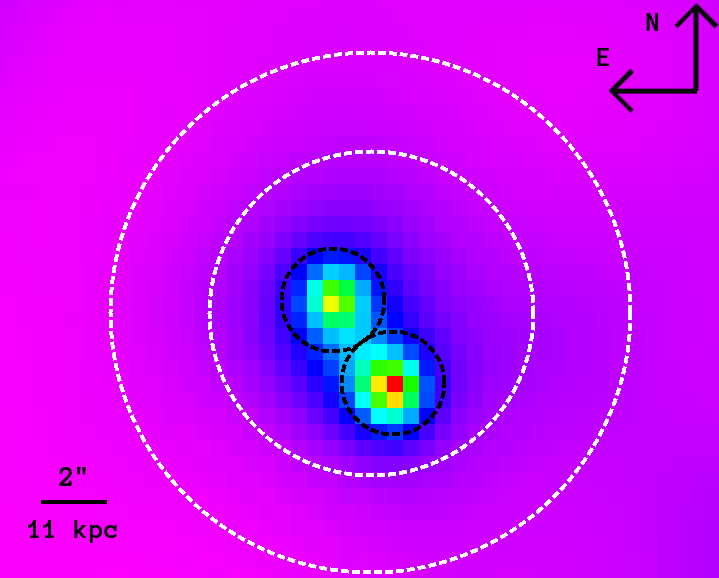}
    \caption{Stacked Chandra ACIS-I image. The black circles are source extraction regions with radii of 1\farcs6. The dotted white annulus is the background extraction region.}
    \label{fig:figure1}
\end{figure}

SDSSJ12 was serendipitously observed with Chandra ACIS-I detector on four occasions in March 2014, with a total exposure time of 188.8 kiloseconds (PI: Young). For data analysis, we used \textsc{CIAO 4.12}, an interactive software package for observation analysis developed by the Chandra X-ray Center \citep{2006SPIE.6270E..1VF}. We used \textsc{CALDB 4.9.3}, the calibration database that stores and provides access to all calibration files required for standard processing and analysis. Observations were reprocessed using the CIAO script \textsc{chandra\_repro}. We merged all observations and created the PSF maps using the \textsc{merge\_obs} script. We calculated that the 90$\%$ time-exposure weighted point-spread function is $\sim$1\farcs6. X-ray spectral fitting was performed using \textsc{SHERPA}, CIAO's fitting and modeling application \citep{2001SPIE.4477...76F}. The source extraction regions were circular apertures with a radius of 1\farcs6, while the background extraction region was an annulus with inner and outer radii of 5$\arcsec$ and 8$\arcsec$, respectively. The extraction regions are shown in Figure \ref{fig:figure1}. Astrometric corrections were performed by running the CIAO tool \textsc{wavdetect} across the full field of view of the X-ray image to detect X-ray sources. We then matched Chandra sources with the USNO-A2.0 catalog \citep{1998yCat.1252....0M}. The CIAO script \textsc{wcs\_match} was then used to take in previously selected X-ray and USNO-A2.0 sources, crossmatch them, and determine transformation parameters to minimize positional discrepancies. Finally, the CIAO script \textsc{wcs\_update} was used to apply the correction parameters and update the event and aspect solution files. The summary of Chandra X-ray observations used in this work is given in Table \ref{tab1:example_table}. \newline \indent
SDSSJ12 was also serendipitously observed with the CFHT's MegaPrime camera on various occasions in various filters. It was also observed in \textit{u, g, r, i,} and \textit{z} bands during the DESI Legacy Imaging Survey, a combination of the Dark Energy Camera Legacy Survey, the Beijing-Arizona Survey, and the Mayall Legacy Survey \citep{2019AJ....157..168D}. We also used observations taken by the Subaru Supreme Hyper Cam \citep{2018PASJ...70S...1M,2018PASJ...70S...2K,2018PASJ...70...66K,2018PASJ...70S...3F}.
\section{Results}
\subsection{Galaxy properties}
The optical spectrum of SDSSJ12 is not available, thus, no information on spectroscopic redshift exists. SDSS Data Release 12 \citep{2015ApJS..219...12A} provides the photometric redshift of $z_{\textup{phot}}$=0.453$\pm$0.086. The DESI Legacy Survey Data Release 8 \citep{2022MNRAS.512.3662D} provides a similar photometric redshift of $z_{\textup{phot}}$=0.435$\pm$0.047. Having two independent surveys provide the same (within the uncertainty) well-constrained photometric redshift is a strong indicator of SDSSJ12's actual distance. We adopt the DESI $z_{\textup{phot}}$ value as a working value. At this redshift, 1$\arcsec$ corresponds to $\sim$5.5 kpc. \newline \indent
We estimated the stellar mass of SDSSJ12 and SDSSJ12b by using the K-corrected g-band mass-to-light ratios and g\--r colors. This approach was adopted from \cite{2003ApJS..149..289B} and was commonly used in the field of dwarf-related dual AGN (e.g., \cite{Reines_2014,Liu_2018}). The relation between the galaxy's stellar mass, absolute magnitudes, and colors was derived from the empirical relation for stellar mass-to-light ratio and is given as: 
\begin{equation}
    \textup{log}(\frac{M_*}{L_g})=a_\lambda+b_\lambda \times color
\end{equation}
where mass and luminosity are expressed in solar units, a$_\lambda$=\--0.499, b$_\lambda$=1.519, color is g\--r color after applying the K-correction, and the Sun's absolute g-band magnitude, M$_{g,\odot}$=5.12 \citep{2007AJ....133..734B}, was used to obtain the g-band luminosity. We analyzed the DESI Legacy Survey g- and r-band images, and we extracted the background-corrected linear fluxes in the units of nanomaggies, for both galaxies, using the 1\farcs12 and 0\farcs96 circular apertures, for g- and r-band images, respectively \citep{Abbott_2018}. We then converted these values into K-corrected g- and r-band absolute magnitudes, adopting the redshift of $z$=0.435 for both galaxies. The resulting stellar masses for SDSSJ12 and SDSSJ12b, conservatively assuming a 50\% uncertainty for the mass-to-light ratio \citep{Secrest_2017}, are  log$(M_*/M_{\odot})$=10.4$^{+0.2}_{-0.3}$ and log$(M_*/M_{\odot})$=9.3$^{+0.2}_{-0.3}$, respectively. These results imply that the merger in question is a minor, $\sim$11:1 mass ratio, merger. Also, the usual upper limit for the dwarf galaxy stellar mass is log$(M_*/M_{\odot})$=9.5 (e.g., \cite{2018MNRAS.478.2576M}) suggesting that SDSSJ12b is an intriguing dwarf-galaxy candidate.
\begin{table}
\begin{centering}
	\caption{Summary of X-ray spectral fitting procedure. }
	\label{tab:tab2}
	\begin{tabular}{ccccc} 
		\hline
		ID$^a$&Cstat/d.o.f.$^b$&N$_\textup{H}$$^c$&F$_X$$^d$&L$_X$$^e$\\
		\hline
		SDSSJ12 & 16.5/13& 2.2$\pm$1.6 & 2.8$^{+0.4}_{\--0.6}$&2.0$^{+0.2}_{\--0.5}$\\
		SDSSJ12b & 10.6/14 & $<$1.4& 3.0$^{+0.3}_{\--0.7}$&2.1$^{+0.3}_{\--0.5}$\\    
		\hline
       \end{tabular}
       \tablecomments{(a) is the name of the galaxy; (b) is the Cstatistics of the best fit and the number of degrees of freedom; (c) is the intrinsic absorption in the units of 10$^{22}$ cm$^{-2}$; (d) is the unabsorbed 0.3-10 keV flux in the units of 10$^{-15}$ erg s$^{-1}$ cm$^{-2}$; (e) is the unabsorbed 0.3-10 keV luminosity in the units of 10$^{42}$ erg s$^{-1}$. The uncertainties are estimated at the 68$\%$ confidence level. The spectral index $\Gamma$ was kept fixed at 1.9.}
\end{centering}       
\end{table}
\subsection{X-ray spectral fitting}
SDSSJ12 and SDSSJ12b were detected with Chandra with 26$\pm$5.5 and 33$\pm$6 net counts, respectively. We used Sherpa to perform the X-ray spectral fitting. The
CIAO tool \textsc{specextract} was used to prepare the auxiliary response file and the redistribution
matrix file for each observation. Then, the \textsc{combine\_spectra} tool was used to create summed spectral files. Due to the low-count nature of both of our sources, we used Cstat statistics \citep{1979ApJ...228..939C} to perform our analysis since it is usually adopted in the literature for low-count spectral fitting. We specified a model
composed of two components: a photoelectric absorption model and a power-law emission model, xszphabs*powlaw1d. The fit is performed by fixing the photon index to a
typical AGN value of $\Gamma$=1.9, and the intrinsic absorption N$_\textup{H}$ was left free
to vary. Then, the fit is repeated leaving both parameters free to
vary, and we check if the fit is significantly improved, (i.e., if
Cstatold-Cstatnew $>$2.71, where Cstatold and Cstatnew are the Cstat values for the fit without and with $\Gamma$ free to vary, respectively). This procedure is adopted from \cite{2018MNRAS.478.2576M} and references therein. Both sources showed better statistics for fits performed with fixed $\Gamma$ value. The resulting unabsorbed 0.3-10 keV luminosities for SDSSJ12 and SDSSJ12b are 2.0$^{+0.2}_{\--0.5}$$\times$10$^{42}$ erg s$^{-1}$ and 2.1$^{+0.3}_{\--0.5}$$\times$10$^{42}$ erg s$^{-1}$, respectively. The summary of the spectral fitting results is given in Table \ref{tab:tab2}. 
\section{Discussion}
\begin{figure*}
	\includegraphics[width=\textwidth,height=10cm]{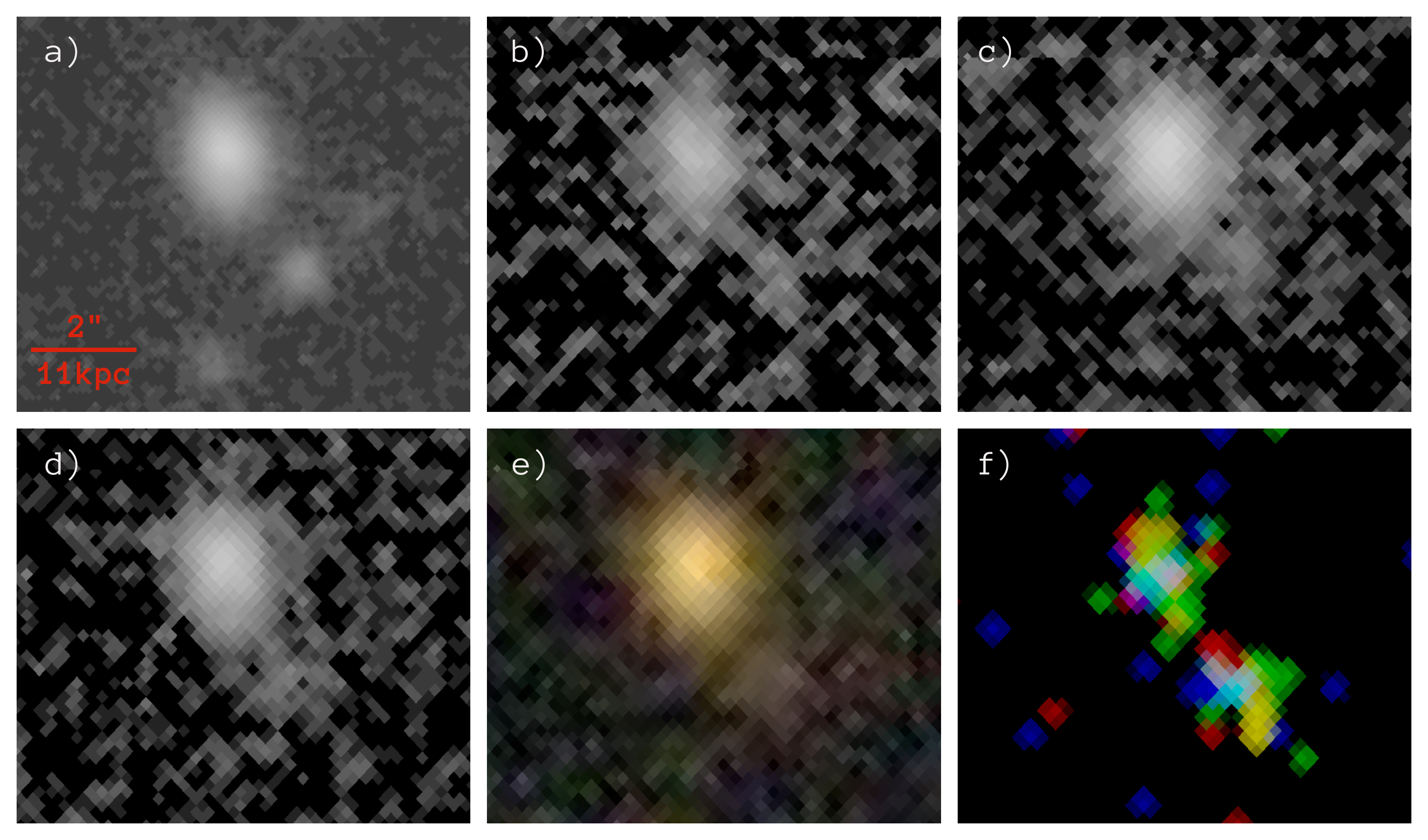}
    \caption{a) Subaru Hyper-Suprime Cam r-band image of SDSSJ12 and SDSSJ12b; b) DESI Legacy Survey g-band; c) DESI Legacy Survey i-band; d) DESI Legacy Survey r-band; e) DESI Legacy Survey color image (g, r, i, z-bands); f) Chandra HiPS image (red-soft band photons, green-medium band photons, blue-hard band photons). The length scale is shown in panel a) and is the same in all other panels.}
    \label{fig2:figure2}
\end{figure*}
\subsection{The stage of the merger}
\begin{figure}
	\includegraphics[width=\columnwidth,height=7cm]{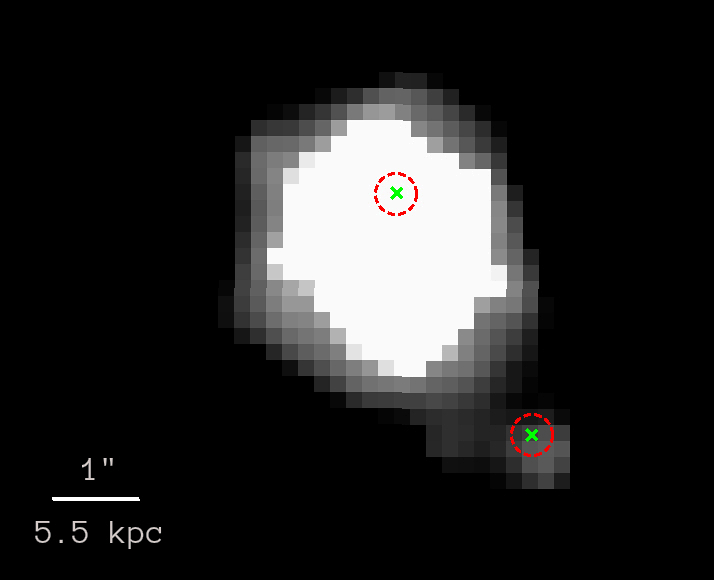}
    \caption{Adaptivelly smoothed CFHT g-band image. Green crosses indicate the positions of X-ray sources with coordinates provided by the CIAO tool \textsc{wavdetect}. The dashed red circles indicate 0$\farcs$24 count contribution weighted positional uncertainties, at the 95\% confidence level \citep{Kim2007}.}
    \label{fig3:figure3}
\end{figure}

Based on this analysis, the two AGN have comparable X-ray luminosities, with the secondary AGN being more likely to be more luminous. The projected separation between the two AGN is 3.2$\arcsec$ or $\sim$17 kpc. The available archival optical imaging suggests that the two galaxies are in the early stages of a merger. The Subaru Suprime Hyper Cam shows two separate nuclei and a faint diffuse emission between them. The DESI Legacy Survey reveals the link between the galaxies more conclusively; images in all available bands indicate the existence of a physical bridge between SDSJ12 and SDSSJ12b. Archival CFHT imaging further solidifies the early-merging hypothesis since it also reveals two separate nuclei and a faint bridge between the two. Various optical images are shown in Figure \ref{fig2:figure2}. Figure \ref{fig3:figure3} shows the adaptively smoothed CFHT image with superposed locations of X-ray sources with positional uncertainties. This physical link between the two galaxies is likely a tidal bridge made out of the material stripped from the smaller galaxy by the more massive companion. The galaxies are likely in the early stages of the merger and similar tidal bridges have been observed in many similar early-stage galaxy interactions (e.g., \cite{Kimbro_2021,Mićić_2023}).
\subsection{AGN activity}
The stellar mass analysis suggests an $\sim$11:1 mass ratio minor merger. As described above, hydrodynamical simulations claim that the secondary black hole is not expected to accrete efficiently during the majority of the minor merger, but it is expected to undergo two phases of growth, one in the early and the other in the late stages of the merger. During the early stage of growth, the secondary black hole is expected to accrete at high Eddington rates, and this would explain the existence of a comparably or more luminous AGN in a dwarf satellite galaxy, powered by a many times smaller black hole. However, this phase is expected to end abruptly once the gas is tidally stripped from the dwarf galaxy. \newline \indent
We calculate the secondary black hole mass, assuming it accretes at the Eddington limit. The Eddington luminosity, in the units of erg s$^{-1}$ is given as:
\begin{equation}
    L_{\textup{Edd}}=1.38\times10^{38}\frac{M_{\textup{BH}}}{M_{\odot}}.
\end{equation}
Assuming the standard bolometric correction of L$_{\textup{bol}}$/L$_\textup{X}$=10 \citep{bolrefId0}, and equating the Eddington luminosity with the bolometric luminosity we find the lower limit of black hole mass to be M$_{\textup{BH}}>$150,000 M$_{\odot}$.
\newline \indent
The purported high accretion rates of SDSSJ12b should be reflected in its X-ray spectral properties. \cite{br2013MNRAS.433.2485B,Brightman_2016} found a statistically significant relation between the accretion rates and the X-ray spectral index $\Gamma$. Higher accretion rates will produce a softer X-ray spectrum with a higher value of $\Gamma$ dominated by the low-energy X-ray photons. On the other hand, a lower accretion rate will typically result in lower $\Gamma$ values and a harder X-ray spectrum dominated by high-energy photons. They concluded that the accretion rate is what drives the conditions of the X-ray corona responsible for X-ray emission and that there is no relation between $\Gamma$ and black hole mass. However, our X-ray sources have low count rates and the spectral fitting was performed using the fixed value of $\Gamma$=1.9, for both sources. The hardness ratio given as $\frac{H-S}{H+S}$, where H and S are net counts in the hard and soft band, can be used as a proxy for the X-ray spectral index $\Gamma$. For that purpose, we filtered the Chandra images by energies in soft (0.3-1 keV) and hard (1-6 eV) bands. After examining the energy-dependent count distribution we note that SDSSJ12 shows no emission under 1 keV, with a 1-$\sigma$ upper limit of $<$1.8 counts \citep{1986ApJ...303..336G}. On the other hand, $\sim$20$\%$ of SDSSJ12b photons have energies under 1 keV, with 6$\pm$2.5 counts. The resulting hardness ratios for SDSSJ12 and SDSSJ12b are HR$>$0.9 and HR=0.6$\pm$0.3, respectively. This hints that SDSSJ12b does indeed have a softer X-ray spectrum, higher spectral index $\Gamma$, and higher accretion rates. However, we note that the hardness ratio difference could also be caused by the higher absorption around the primary AGN SDSSJ12. Even though not conclusively, this tentatively corroborates the scenario of a dwarf-related early-stage minor merger, with an AGN powered by a low-mass black hole, accreting at high Eddington rates, until the activity abruptly shuts down due to tidal stripping.
\subsection{Comparison to Was49}
SDSSJ12 is the second-ever minor merger with a comparably/more luminous AGN residing in a secondary nucleus after Was49. However, unlike SDSSJ12, Was49 is a late-stage merger and the secondary nucleus has already circularized within the primary galaxy disk. Was49 was first identified by \cite{1983ApJ...272...68W} as a galaxy with two knots of emission. \cite{both1989ApJS...70..271B} and \cite{moran1992AJ....104..990M} performed follow-up spectroscopic studies and found that the primary galaxy, Was49a is a Seyfert 2 galaxy, and the secondary galaxy Was49b a Seyfert 1.8 galaxy. They also found that the secondary Was49b has an unusually high bolometric luminosity of L$_{\textup{bol}}>$10$^{45}$ erg s$^{-1}$, significantly stronger than that of the primary Was49a. \cite{Secrest_2017} studied the Was49 system using X-rays and confirmed that the secondary nucleus bolometric luminosity exceeds 10$^{45}$ erg s$^{-1}$. However, the primary nucleus is detected with only seven X-ray counts, all with energies under 2 keV, suggesting strong contamination by non-AGN photons. With the data they had in hand, they were unable to further examine the nature of the primary AGN in Was49a. Finally, they found that the secondary AGN is powered by an overmassive black hole, M$_{\textup{BH}}$=1.9$\times$10$^8$ M$_{\odot}$, accounting for 2.3\% of the stellar mass of the Was49b nucleus. It was suggested that this unusually high ratio is due to a significant loss of stellar material due to interaction with Was49a. On the other hand, the SDSSJ12 system is an early-stage merger, with two separate nuclei, and an unambiguous presence of dual AGN. Even though the accretion rates of the secondary AGN during an early-stage merger are expected to approach Eddington limits, SDSSJ12b is $\sim$two orders of magnitude fainter than Was49b. This suggests, unlike the overmassive black hole case of Was49b, a scenario of a low-mass black hole powering the secondary AGN.
\section{Conclusions}
In this paper, we presented the evidence for an ongoing merger between the galaxy SDSSJ12 and its smaller companion SDSSJ12b at $z$=0.435. The available archival imaging reveals a physical link in the form of a tidal bridge between the two galaxies, suggesting they are in the early stages of a merger. We analyzed Chandra X-ray data and showed that two luminous point X-ray sources reside in each nucleus. The primary and secondary AGN have comparable X-ray luminosities, with the secondary being more likely to be more luminous, making this the second discovery ever of this kind, and the first in an early-stage merger. Further analysis suggests that the mass ratio is 11:1 and that the secondary AGN is powered by a low-mass black hole undergoing high accretion rates. This is in accordance with computational predictions, suggesting that the secondary AGN in minor mergers undergoes an episode of high accretion during the early stages of the merger until it abruptly shuts down due to the tidal stripping of its gas. The follow-up multiwavelength observations will shed more light on this object, allowing us to study the, mostly inaccessible, accretion physics of the secondary AGN in minor mergers and rapid black hole growth mechanisms.
\paragraph{Acknowledgements}
We thank an anonymous referee for helpful and insightful comments. MM is supported by the University of Alabama Graduate Council Fellowship. BNW and OJH are supported by the University of Alabama Department of Physics and Astronomy Undergraduate Barr Scholarship. Authors acknowledge the use of the \href{https://www.astro.ucla.edu/~wright/CosmoCalc.html}{Cosmology Calculator} \citep{2006PASP..118.1711W} and the \href{http://kcor.sai.msu.ru/about/}{K-correction calculator} \citep{2010MNRAS.405.1409C}. Authors acknowledge the use of observations obtained with MegaPrime/MegaCam, a joint project of CFHT and CEA/DAPNIA, at the Canada-France-Hawaii Telescope (CFHT) which is operated by the National Research Council (NRC) of Canada, the Institut National des Sciences de l'Univers of the Centre National de la Recherche Scientifique of France, and the University of Hawaii.

\bibliographystyle{mnras}
\bibliography{main}
\end{document}